# Open Access repositories and journals for visibility: Implications for Malaysian libraries

**A.N. Zainab**
Digital Library Research Group,
Faculty of Computer Science & Information Technology,
University of Malaya, MALAYSIA
e-mail: zainab@um.edu.my

## ABSTRACT

*This paper describes the growth of Open Access (OA) repositories and journals as reported by monitoring initiatives such as ROAR (Registry of Open Access Repositories), Open DOAR (Open Directory of Open Access Repositories), DOAJ (Directory of Open Access Journals), Directory of Web Ranking of World Repositories by the Cybermetrics Laboratory in Spain and published literature. The performance of Malaysian OA repositories and journals is highlighted. The strength of OA channels in increasing visibility and citations are evidenced by research findings. It is proposed that libraries champion OA initiatives by making university or institutional governance aware; encouraging institutional journal publishers to adopt OA platform; collaborating with research groups to jumpstart OA institutional initiatives and to embed OA awareness into user and researcher education programmes. By actively involved, libraries will be free of permission, licensing and archiving barriers usually imposed in traditional publishing situation.*

## INTRODUCTION

E-repositories and journals supported by open source software mainly contain Open Access literature. Open Access literatures are "digital, online, free of charge and free of most copyright and licensing restrictions" (Suber 2004, 2007). There are two ways in which Open Access literature can be delivered, and these are through: (a) Open Access repositories and (b) Open Access journals. There is a distinct difference between the two. The former can be either hosted by single institutions or cross institutional and provides an avenue for individuals and institutional members to deposit works which are then made freely available. In this situation, no peer reviewing process is undertaken as the repository merely acts as an archival platform. The contents are mainly preprints, post prints or both, unpublished scholarly works such as theses, dissertations, final year project reports, research reports and teaching resources. These contents are usually maintained by universities, research laboratories or groups, professional societies and associations that commission the repositories. The latter, Open Access (OA) journals refer to electronic journals, which give access to all users and are subscription free. Peer reviewing is undertaken in OA journals and, in this case the accepted articles will then be made freely available to users. There is a difference between Open Access and free access. Open Access imply free to view, use, distribute and the copyright is held by the author. Free access can mean free access but with restrictions in terms of use, to redistribute and the copyright is often held by the publishers or creators.

The Berlin Declaration (2003) promotes the Internet as the functional instrument for worldwide sharing of scientific knowledge derived from research funded by educational institutions, libraries, archives and museums. The Declaration therefore called for an open committment to Open Access to contributions of all forms of knowledge. As a result of the Open Access movements, there is now a number of open source software available for building both repositories and journals. Most of the Open Access repositories and journals are free. In some cases, the cost is incurred by the authors themselves for publishing their works, whilst the works are completely free to users. An Open Source Software Directory is available (http://www.opensourcesoftware directory.com/), developed by Jeroen Verhoectx, which currently lists 843 applications divided into four categories; for home users, businesses, administrators and developers.

## OBJECTIVES

The objectives of this paper are:
- (a) to define and describe the growth trends of Open Access repositories and Open Access journals and
- (b) to propose the roles libraries can play in promoting these initiatives.

## METHODOLOGY

This is a descriptive paper where information about the trends and growth of repositories and journals are collated from monitoring initiatives comprising *ROAR* (*Registry of Open Access Repositories*), *Open DOAR* (*Open Directory of  Repositories*), *DOAJ* (*Directory of Journals*) and *Web Ranking of World Repositories* by the Cybermetrics Laboratory in Spain. The information about increased visibility and citations achieved by articles are evidenced from published literature.

## REPOSITORIES

Many of these repositories are set up by universities or research institutions to handle their own institutional research resources. *D-Space,* for example, was set up by Massachusetts Institute of Technology to hold their entire intellectual output. This repository is linked to similar archives at other research institutions, thus creating a "seamless worldwide network, where multiple databases could be searched as if they were a single entity" (Ware 2004). *D-Space* initially costs MIT US$2.4 million and was jointly sponsored by Hewlett-Packard of Palo Alto, California. Such a repository is vital in helping institutions create and maintain their own archive for the posterity of all digital documents and data they generate themselves. This is different from cross-institutional repositories such the *arXiv.org* (http://arxiv.org/), an archive designed to serve communities in specific disciplines (physics, mathematics, non-linear sciences, computer science, quantitative biology, quantitative finance, and statistics). The institutional repositories provide publishing tools, which could be easily handled by academics to self archive their own works. Most archives comply with the standards initiated by the Open Archive Initiatives (OAI) (http://www.openarchives.org/) for describing documents and digital objects. The

standard unite all distributed archives that use it and facilitate searching as if they were one, by either using the search engine provided by the OAI service providers or by general search engines like Google. These repositories allow access to most of their content with some imposing restrictions to documents such as theses and dissertations or e-textbooks written by their academics. Such repository is becoming a common feature of a modern university and an indication of a reformed scholarly communication where the institution provides a set of services for the management and dissemination of digital materials created by the institution and its members. As these resources are on Open Access, it would enhance institutional prestige by making their research output more visible. This is therefore done in the spirit which Harnad (2003) had previously advocated, that is, "self-archiving". A list of institutional archives worth mentioning is listed in *SPARC: collected repositories* (SPARC 2007-2010) and is available at http://www.arl.org/sparc/repositories/collectedrep.shtml, the *Registry of Open Access Repositories (ROAR)* (Registry 2007-2010) and *Open DOAR* (Open DOAR 2006-2010).

Even though the number of OA repositories is growing, the idea of institutional self-archiving has only caught on in recent years as there was concern about plagiarism. Ware (2004) surveyed 45 institutional repositories and found that the average number of contents held was low. Most of the contents were pre-prints, theses and dissertations submitted by early ICT adopters.  This situation was found to be the result of poor academic participation and this reluctance cannot be explained as Gadd, Oppenheim and Probets (2003) and Crow (2002) found that journal publishers did allow self-archiving.

Repositories can either be institutional or cross-institutional or discipline-based. There are many cross-institutional repositories which hold e-print and post-prints (Hitchcoc, 2003), but an example of a successful venture is the *arXiv.org* (http://arxiv.org/). This repository was developed by Ginsparg, Paul (1996) at the Los Alamos National Laboratory in 1991 but has since moved to Cornell University and funded by Cornell, the National Science Foundation and participating institutions. This archive focuses on research papers in physics and its related disciplines, nonlinear science, mathematics, computer science and quantitative biology submitted by researchers from all over the world. This archive pays attention to the needs of users and authors and plays down the role of the publisher as processes are highly automated. Users can retrieve papers from the archive either through an online web interface, or via e-mail links. Similarly, authors can submit their papers or reports to the archive, by either using the web interface, ftp or using their e-mail. Authors can update their submissions if they choose to, and previous versions of articles remain available for users to view. Users can also register to automatically receive a listing of newly submitted papers in areas of interest to them. An example of a domain well covered by this archive is high energy physics theory (http://arxiv.org/hep-th). The archive was started in 1991 and was intended for less than 200 physicists working on "string theory". Within a few months, users of the archive grew to over 1000 and by 2009 it typically processed 489,368 transactions per day (Ginspar, 1994; 2002). As in 2010, the repository held roughly 614,672 full-text e-prints and growth rate of more than 40,000 new submissions per year. Its usage grew because physicists need to communicate quickly and easily. This type of publication channel soon becomes indispensable to physicists, especially for those in developing countries. The repository works on the simple premise that if researchers are writing without the expectation of making money directly from their efforts, then there would be no reason why anyone else should. Brinkman (2002) remarked that physicists who used the *arXiv* site did not appear concerned that the papers on it were not refereed. To the physicists the repository acts as a comprehensive "archival aggregator", a place where they could browse or search and be assured that the relevant articles they need can be found and if not, it is because it does not exist. This model works very well for the physicists but are slow to take off in a field such as medicine, where posted materials are substantially reviewed before they are published in an archive (Kling, Spector and McKim 2002) for the simple reason that wrong reporting may result in deaths! *The New England Journal of Medicine* has indicated that they do not accept preprint submissions. A list of clinical medicine journals that will (29 titles) and will not (21 titles) accept preprints appears at http://clinmed.netprints. org/misc/ policies.shtml.

Most repositories are dedicated to the science and technology disciplines. Besides the *arXiv.org*, other equally well known archives are :

> *CERN* document server (http://preprints.cern.ch), which provide full text coverage of preprints, articles, books, journals and photographs since 1994 and include links to their preprint servers in the subjects of psychology, neuroscience, linguistics and cognitive sciences;
>
> *Clinmed Net prints* (http://clinmed. netprint.org), which is produced by the British Medical Journal and Highwire Press, providing a place where authors could archive completed studies and original research preprints;
>
> *PubMedCentral* (http://www.pubmedcentral. nih.gov/) published by the US National Library of Medicine's digital archive of life sciences journal literature, and includes full-text articles, data tables, streaming videos and high resolution images;
>
> *Highwire* initiative at Stanford University;
>
> Examples of e-print archives in the arts and social sciences are *eScholarship* repository (University of California' digital repository of humanities and social science research);
>
> *Social Science Research Network* (SSRN, providing access to over 30,800 papers and over 49,200 abstracts;
>
> *Educationon-line* available at http://www.leeds. ac.uk/educol/, providing free access to conference papers, working paper, preprints;
>
> *PhilSci Archive* (philosophy of science, hosted by the Department of Philosophy and of History, University of Pittsburgh and available at http://philsci-archive.piutt.edu/); *Preprints on conservation laws* (administered by the Department of Mathematical Sciences, Norwegian University of Science and Technology at Trondheim since 1996); *RePEc* (Repository on Economics at http://repec.org, provide access to over 177,000 records and over 86,000 are available online);
>
> *WoPEc archives* (Working papers in Economics, the economic network database of working papers, containing over 80,000 documents, 53,035 working papers and 41,895 journal articles and available at http://netec.mcc. ac.uk/WoPEc/data/paper Series.html.; An example of a cross-institutional repository in Malaysia is MyAIS (hhtp://myais.fsktm.um.edu.my) and MyManuskrip (http://myManuskrip.fsktm. um.edu.my). The former archives articles from Malaysian Scholarly journals and the latter archives manuscripts from libraries at the University of Malaya and *Dewan Bahasa dan Pustaka.*

## Monitoring

There are two initiatives that monitor the existence of repositories world-wide. The first is *ROAR* or *Registry of Open Access Repository* (Registry 2005-2010) developed by Tim Brody at the University of Southampton, UK. ROAR listed 1606 repositories world –wide and has a useful option where repositories can submit their domain to be included in the registry. ROAR listed 19 Malaysian repositories. The second is *Open DOAR* (*Open Directory of*

*Repository)* hosted by the University of Nottingham in UK which makes public their services since 2006. This service has been awarded the SPARC Europe Award for Outstanding Achievements in Scholarly Communications. *Open DOAR* (2006-2010) listed 1650 repositories world-wide and Malaysia is listed as having 11 repositories (0.6%) and this is quite encouraging compared to other ASEAN countries (Philippines 1, Singapore 2, Thailand 3, Indonesia 6, Vietnam 1). Obviously, the larger Asian countries lead such as India (41) and Japan (79) and Taiwan (35). Figure 1 shows the world-wide regional distribution and Table 1 shows the distribution in Asia.

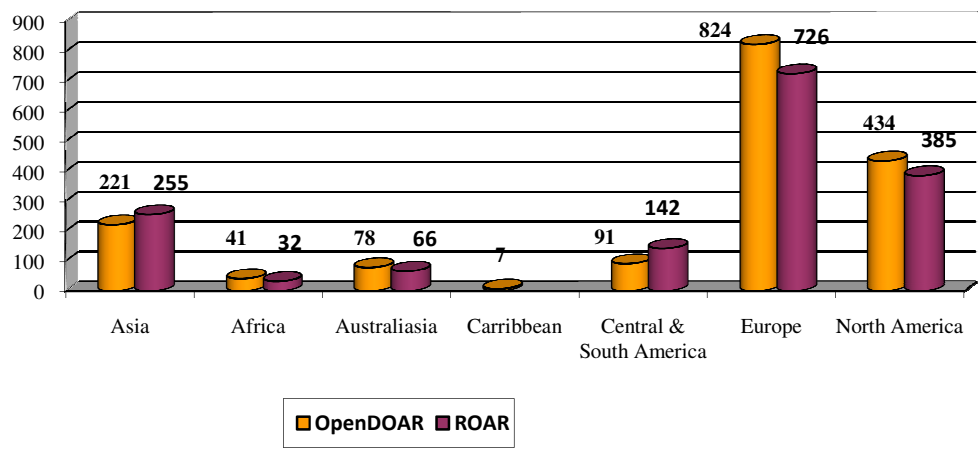

Figure 1: Repositories World Wide Based on Open DOAR (n=1696) and ROAR (n=1606)

Table 1: Distribution of Repositories in Asia Listed in ROAR and Open DOAR

| Countries in Asia | ROAR (n=255) | OpenDOAR (n=221) |
|---|---|---|
| Japan | 78 | 79 |
| India | 53 | 40 |
| Taiwan | 45 | 35 |
| Malaysia | 17 | 11 |
| Turkey | 12 | 9 |
| China | 12 | 9 |
| Indonesia | 10 | 6 |
| S. Korea | 4 | 5 |
| Hong Kong | 4 | 0 |
| Iran | 3 | 2 |
| Saudi Arabia | 1 | 3 |
| Azerbaijan | 2 | 1 |
| Bangladesh | 2 | 2 |
| Kyrgyzstan | 2 | 2 |
| Singapore | 2 | 2 |
| Sri Lanka | 2 | 1 |
| Thailand | 2 | 3 |
| Israel | 1 | 2 |
| Afghanistan | 0 | 1 |
| Georgia | 0 | 1 |
| Kazakhstan | 1 | 1 |
| Philippines | 1 | 1 |
| Pakistan | 1 | 2 |
| Nepal | 0 | 1 |
| Qatar | 0 | 1 |
| Vietnam | 0 | 1 |

The eleven Malaysian repositories listed in *Open DOAR* are shown in asterisk in Table 2. The rest are repositories reported in the *Registry of Open Access Repository* (ROAR) (Registry 2005-2010).

Table 2: Institutional Repositories in Malaysia as at 5.7.2010 (n=20)

| Repositories | Software | Contents | Access |
|---|---|---|---|
| • DSpace@UM<br>Digital Library group, Faculty of Computer Science & Information Technology, University of Malaya http://mymanuskrip.fsktm.um.edu.my/Greenstone/cgi-bin/library.exe | DSpace | Master dissertations, Ph.D theses, Final year project reports 763 items (6-7-2010) | • Unrestricted, full access to masters and Ph.D materials except final year project report. |
| • DSpace/Manakin Repository<br>Universiti Tenaga Nasional Library http://dspace.uniten.edu.my/xmlui/ | DSpace | Articles, Conference papers, Digital images, In house publications, Manuals, news articles, examination papers, project papers, Research reports, Theses & dissertations, Speeches. 453 items | • Restrictions for examination papers & theses. |
| • Elmtiyaz@Usim Intellectual/Manakin Repository<br>Universiti Sains Islam Malaysia http://ddms.usim.edu.my/ | Dspace | 2694 items. Mainly theses and dissertations. Also include academic project papers, conference papers, examination papers, news clippings, research reports. | • Full access to items |
| • EPrints@USM<br>Universiti Sains Malaysia http://eprints.usm.my/cgi/oai2 | Eprint3 | Journal articles, Conference papers, theses.<br>17,611 items (5-7-2010) article - 227, books - 88, book section - 56, conferences − 592, images - 83, monographs − 490, others − 156, teaching resources − 15050, theses − 919) | • Full access to most contents except theses.<br>• Books are uploaded in a single file, therefore slow to download − suggest break book contents into chapters<br>• Restricted access to theses, title, contents pages, chapter 1 up to about 30 - 50 pages only. |
| • MyAIS (Malaysian Abstracting and Indexing System)<br>Digital Library Research Group, Faculty of Computer Science and Information Technology, University of Malaya. Available at: http://myais.fsktm.um.edu.my/ | Eprint | Mainly journal articles, conference papers.<br>8,880 items (5-7-2010)<br>Journal articles 8447, Books 16, book chapters 26, conference papers 376, monographs, 5 and theses 10) | • Full text access to most items unless the restrictions is required by copyright owners. |
| • PTSL UKM Repository<br>Tun Seri Lanang Library, Universiti Kebangsaan Malaysia. Available at: http://eprints.ukm.my/<br>UKM Library has indicated that the University has created a repository named E-Repository Penerbitan available at http://smk5.ukm.my/epenerbitan/. However this repository is not on as it requires password to login. | Eprint | Conference papers and journal articles.<br>208 items (5-7-2010) | • Full access to most contents |
| • MyManuskrip.fsktm.um.edu.my<br>Digital Library Research Group, | Greenstone | Malay manuscripts, research papers and reports on manuscript studies 178 items (6-7-2010) | • Full access to all contents |

| | | 69 – DBP collection, 99 UM collection, 4 published items, 6 others | |
|---|---|---|---|
| • SHDL@mmu digital repository<br>Multimedia University Malaysia<br>http://shdl.mmu.edu.my/ | Eprint3 | Journal articles, books, book sections, conferences, monographs, others, theses.<br>1450 items (5-7-2010) | • Full access to articles and selected conference papers<br>• Restricted access to books, book section and theses |
| • UiTM digital repository<br>Universiti Teknologi Mara<br>http://eprints.ptar.uitm.edu.my/ | Eprint3 | Mainly journal articles, conference papers. 288 items (5-7-2010) | • Full access to journal articles |
| • UM Digital Repository (University of Malaya Library)<br>http://eprints.UM.edu.my/cgi/oai2 | Eprint3 | Journal Articles, Conference papers, books<br>1446 (2010-02-04) | • Full access to articles & conference papers.<br>• Restricted access for books access (access only to title and contents pages and bibliographies. |
| • UMP@institutional repository<br>Universiti Malaysia Pahang<br>http://umpir.ump.edu.my/information.html | Eprint3 | Journal articles, conference papers, theses, new clippings, images<br>698 items (5-7-2010)<br>Article 1, book section 1, conference 1, images 37, other 1, theses 667. | • Full access to most contents except theses<br>• Restricted access for theses to 24 pages only |
| • UniMAP Library Digital Repository<br>Universiti Perlis Malaysia<br>http://dspace.unimap.edu.my/dspace/ | Dspace | Conference papers; theses, journal articles, pass examination papers, newspaper clippings.<br>7146 items (16-11-2004) | • Restricted access to all resources<br>• Cannot view fulltext, contents restricted to content pages, abstracts. A small number of journal articles are given full text access |
| • Universiti Putra Malaysia Institutional Repository (PSAS IR)<br>http:/psasir.up,.edu.my/cgi/oai2 | Eprint3 | Journal articles, conference papers. Learning objects. Theses<br>5869 items (5-7-2010)<br>Articles 2159, conferences 40, inaugural lectures 43, newspaper clippings, 756, theses 2512, upm news 359. | • Full access to most articles, conferences, newspaper clippings.<br>• Restricted access to theses (title page, contents pages, abstracts, part of chapter one) – about 25 pages only. |
| • UTHM repository<br>Universiti Tun Hussein Onn<br>http://eprints.uthm.edu.my/ | Eprint3 | Mainly conference papers<br>159 items (5-7-2010) | • Full access to conference papers |
| • Universiti Teknologi Malaysia Institutional Repository<br>http://eprints.utm.my/ | Eprint | Journal articles, conference papers, Theses; Books.<br>7413 items (5-7-2010) | • Full access to most contents |
| • UTEM Perpustakaan Library<br>Universiti Teknikal Malaysia Melaka<br>http://library.utem.edu.my/index.php?option=com_docman&Itemid=208 | | Journal articles, reports, proceedings, theses, final year projects, journal contents, speeches, exam papers<br>2702 items (5-7-2010).<br>Reports 63, proceedings 352, theses 174, final year projects 614, journal contents pages 754, speeches 161, examination paper 986. | • Restricted access. Access to the first 26-27 pages only |
| • UTP Institutional Repository<br>Universiti Teknologi Petronas<br>http://eprints.utp.edu.my/ | Eprint3 | Mainly journal articles and conference papers.<br>1282 items (5-7-2010) | • Restricted access to most contents. |
| • UUM IRepository http://eprints.uum.edu.my/cgi/oai2 | Eprint3 | Mainly theses, includes journal articles, conference papers. 1791 items (5-6-2010) Articles 268, conferences 185, theses 1338. | • Restricted access to theses (title and contents pages, few pages of the first chapter, references, appendices, questionnaire used.<br>• Full access to selected journal articles, conference papers |
| • ethesis@UUM (Electronic theses and dissertations) Universiti Utara Malaysia http://ep3.uum.edu.my/view/subjects/ | Eprint3 | Theses 1706 items (5-7-2010) | • Full access to theses |
| • WorldFish Centre Publications World Fish Centre, Penang http://www.worldfishcenter.org/v2/pubs.html | | Articles; References; Conferences; Unpublished; Books; Special 549 items (2009-05-13) | • Full access to all materials |

The listing above indicates that:
- The repositories do not undertake peer reviewing and provide an archival option to institutional works.
- Most of the Malaysian repositories are institutional or department based. Only two are cross institutional (MyAIS and MyManuskrip)
- Most repositories deposit all types of items; scholarly and non-scholarly, including journal articles, conference papers, examination questions, final year student project reports, theses and dissertations, research reports, images, news clippings and teaching resources.
- UUM and UM have provided access to their theses and dissertation collections in separate repositories, which makes good sense especially if the collections are delivered full-text, which would need huge storage and computing power for speedier access.
- Most theses collections except for UM are delivered in a single PDF file, which is cumbersome for users as downloading and opening the folder will consume more time as some of the files can be as large as over 50MB.
- Some Universities such as UPM and UTEM provide access to only about 25 pages of their theses collections.
- Most repositories are hosted and managed by the libraries. At UM, three repositories are hosted at the Faculty of Computer Science (MyAIS, MyManuskrip and DSpace@UM).

The Berlin Declaration (Berlin 2003) identifies two conditions of OAI:
  (a) Users should be given free access and a license to copy, use, distribute, transmit and display of all contributions by authors and right holders; and
  (b) Academic institutions, scholarly society, governance agency are responsible for making available and maintaining the digitized content of works in repositories which are OAI compliant.

This will ensure that contributions are available for unrestricted access in a long term archiving environment.  In this spirit then, the restrictions imposed by Malaysian repositories need to be looked at seriously as this would mean using systems to behave like a rich library catalogue (providing rich metadata information and limiting access to items). It must be remembered that accessibility will result in usability and hopefully citation. It is a matter of librarians persuading university governance to adopt the policy.

Previous studies have indicated that articles are being cited more (Antelman 2004; Eysenbach 2006). Swan (2010) examined 31 studies about the citation advantage of OA articles and found 27 studies reported positive citation advantages and 4 studies reported no citation advantage. Lawrence (2001) compares citation counts and online availability of 119, 924 conference titles in computer science obtained from DBLP (dblp.uni.ytier.de) using Research Index and exclude self-citation. Lawrence found correlation between the number of times an article is cited and the probability that the articles are freely available online. The mean citation of offline articles was 2.74 compared to 7.03 for OA online articles. Other studies have stressed that the increase in citation of OA articles are discipline dependent, that is citations occur more in fields such as life sciences, engineering, physics and mathematics (Craig et al. 2007). The list of highly ranked repositories in Open DOAR substantiates this point as repositories in the top ten perform well in terms of accessibility and web visibility. Also, the highly ranked repositories are those in the fields where the tradition for self-archiving and using open archive repositories are highly preferred as exemplified by Arxiv.org amongst the physicists.

## JOURNALS

Unlike repositories, journals especially those which are scholarly, are peer reviewed. Nielsen (2010) identifies three types of journal publishing models; the traditional toll access journals, the golden journals and the green journals (see Figure 2).

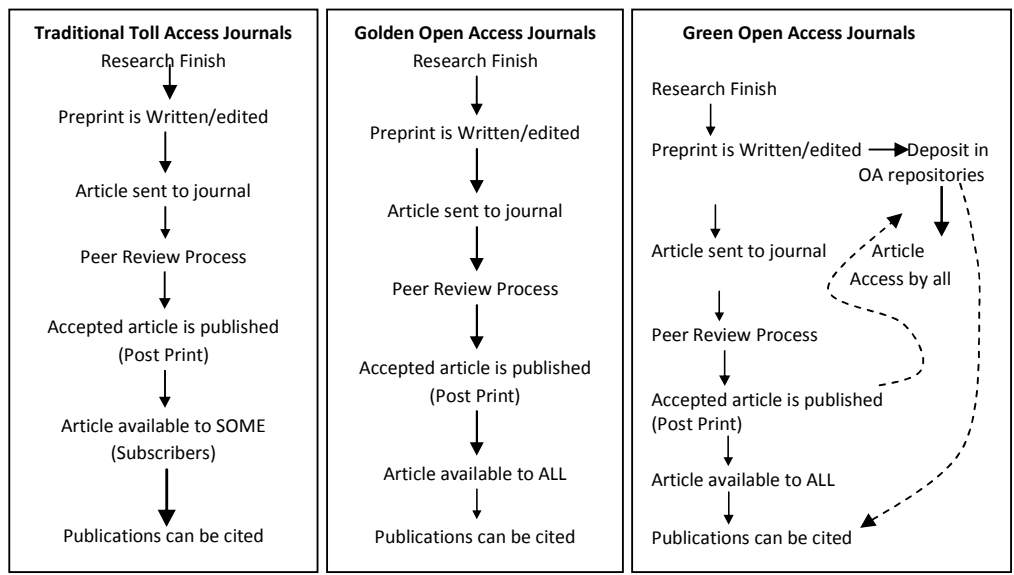

Figure 2: Journal Publishing Process

In the Golden publishing model, the author pay-to-publish practice is often used as is practiced by medical journals such as *PLoS Medicine* which is a peer reviewed medical journal, where the cost of publishing is transferred to authors and users are given full access to articles. In the Green approach, user access is provided at various stages that is:

a) at the pre-print stage, where authors submit to OA repositories to get feedback from readers, improves on their articles before submitting to OA journals which becomes accessible to all;

b) at the accepted stage where, the authors' submissions are peer reviewed before being accepted by OA journals which, subsequently make them available to all, and

c) at the post-print stage where, authors submit their post-prints to an OA repository after informing or obtaining permission from OA journals publisher of his intention to deposit his article in his institutional or cross institutional repositories which makes their works accessible by all. In the "Green" case the visibility is increased

The number of electronic journals is increasing. This growth is derived from institutional and professional publishers who want to increase access to the contents of their journals without any restrictions. The pull factor is the increase in readership and citation to the contents. The *Directory of Journal* (DOAJ) (2010) published by Lund University mooted by the First Nordic Conference on Scholarly Communication in Lund, Copenhagen and initially funded by the Open Society Institute indicates the existence of 5160 journals with contents of 416,421 articles. The directory focused only on journals that provide full text, are peer reviewed and scholarly. Currently, it is estimated that 20% of total articles published are on Open Access (Hitchcock 2004, updated 2010). A total of 31 Malaysian Journals are listed in DOAJ (Table3).

## Table 3: Malaysian OA Journals in DOAJ (n=31)

1. **3L Language, Linguistics and Literature : the Southeast Asian Journal of English Language Studies**
ISSN: 01285157
Subject: Languages and Literatures
Publisher: Penerbit UKM ; Start year: 2003
Added to DOAJ: 2010-04-16

2. **ASEAN Journal of Teaching & Learning in Higher Education**
ISSN: 19855826
Subject: Education
Publisher: National University of Malaysia ; Start year: 2010
Added to DOAJ: 2010-01-25

3. **Asian Academy of Management Journal**
ISSN: 13942603 ; EISSN: 19858280
Subject: Business and Management
Publisher: Universiti Sains Malaysia Press ; Start year: 2002
Added to DOAJ: 2010-03-31

4. **Biomedical Imaging and Intervention Journal**
ISSN: 18235530
Subject: Medicine (General) , physics, radiobiology
Publisher: University of Malaya ; Start year: 2005 ;
Added to DOAJ: 2005-08-31

5. **CFD Letters**
ISSN: 21801363
Subject: General and Civil Engineering
Publisher: ISSR ; Start year: 2009
Added to DOAJ: 2009-12-09

6. **Concrete Research Letters**
ISSN: 21801371
Subject: Construction
Publisher: ISSR ; Start year: 2010
Added to DOAJ: 2010-03-31

17. **Journal of Physical Science**
ISSN: 16753402 ; EISSN: 19858337
Subject: Science (General)
Publisher: Universiti Sains Malaysia Press ; Start year: 2007
Keywords: physics, chemistry, material science
Added to DOAJ: 2010-04-01

18. **Jurnal Kejuruteraan**
ISSN: 19854625
Subject: General and Civil Engineering
Publisher: Penerbit UKM ; Start year 2006
Keywords: technology ; Added to DOAJ: 2010-03-10

19. **Jurnal Kemanusiaan**
ISSN: 16751930
Subject: Business and Management
Publisher: Universiti Teknologi Malaysia ; Start year: 2003
Added to DOAJ: 2010-03-03

20. **Jurnal Pendidikan Malaysia**
ISSN: 21800782
Subject: Education
Publisher: National University of Malaysia ; Start year: 2005
Added to DOAJ: 2009-10-30

21. **KEMANUSIAAN : The Asian Journal of Humanities**
ISSN: 13949330 ; EISSN: 19858353
Subject: Languages and Literatures
Publisher: Universiti Sains Malaysia Press ; Start year: 2008 . Added to DOAJ: 2010-04-01

22. **Malaysian Family Physician**
ISSN: 1985207X ; EISSN: 19852274
Subject: Medicine (General)
Publisher: Academy of Family Physicians of Malaysia
Start year: 2006 ; Added to DOAJ: 2008-01-15

7. **Elektrika: Journal of Electrical Engineering**
ISSN: 01284428
Subject: Electrical and Nuclear Engineering
Publisher: University Teknologi Malaysia
Start year: 2006 ; Added to DOAJ: 2008-01-24

8. **GEMA Online Journal of Language Studies**
ISSN: 16758021
Subject: Languages and Literatures --- Linguistics
Publisher: Universiti Kebangsaan Malaysia ; Start year: 2001
Added to DOAJ: 2007-07-31

9. **International Journal of Asia-Pacific studies**
ISSN: 18236243
Subject: Multidisciplinary
Publisher: USM Press ; Start year: 2005
Added to DOAJ: 2006-09-27

10. **International Journal of Biometric and Bioinformatics**
ISSN: 19852347
Subject: Biology --- Mathematics
Publisher: Computer Science Journals ; Start year: 2007
Added to DOAJ: 2009-06-02

11. **International Journal of Computer Science and Security**
ISSN: 19851553
Subject: Computer Science
Publisher: Computer Science Journals ; Start year: 2007
Added to DOAJ: 2009-06-02

12. **International Journal of Engineering**
ISSN: 19852312
Subject: General and Civil Engineering
Publisher: Computer Science Journals ; Start year: 2007
Added to DOAJ: 2009-05-20

13. **International Journal of Image Processing**
ISSN: 19852304
Subject: General and Civil Engineering
Publisher: Computer Science Journals ; Start year: 2007
Added to DOAJ: 2009-06-02

14. **International Journal of Security**
ISSN: 19852320
Subject: Computer Science
Publisher: Computer Science Journals ; Start year: 2007
Added to DOAJ: 2009-06-02

15. **Jebat : Malaysian Journal of History, Politics and Strategic Studies**
ISSN: 01265644 ; EISSN: 21800251
Subject: History
Publisher: Universiti Kebangsaan Malaysia ; Start year: 2007
Added to DOAJ: 2010-04-14

16. **Journal of Construction in Developing Countries**
ISSN: 18236499 ; EISSN: 19858329
Subject: Construction
Publisher: Universiti Sains Malaysia ; Start year: 2006
Added to DOAJ: 2010-04-16 11:15:35

23. **Malaysian Journal of Community Health**
ISSN: 16751663
Subject: Public Health
Publisher: Universiti Kebangsaan Malaysia ; Start year: 2006
Added to DOAJ: 2010-01-1

24. **Malaysian Journal of Medical Sciences**
ISSN: 1394195X
Subject: Medicine (General)
Publisher: Universiti Sains Malaysia ; Start year: 2002
Added to DOAJ: 2007-11-20

25. **Malaysian Journal of Pharmaceutical Sciences**
ISSN: 16757319 ; EISSN: 19858396
Subject: Therapeutics
Publisher: Universiti Sains Malaysia ; Start year: 2004
Added to DOAJ: 2010-04-26

26. **Matematika**
ISSN: 01278274
Subject: Mathematics
Publisher: Universiti Teknologi Malaysia ; Start year: 1997
Added to DOAJ: 2008-08-13

27. **Neurological Journal of South East Asia**
ISSN: 1394780X
Subject: Neurology
Publisher: ASEAN Neurological Association ; Start year: 1996. End year: 2003 Continued by Neurology Asia
Added to DOAJ: 2007-02-22

28. **Neurology Asia**
ISSN: 18236138
Subject: Neurology
Publisher: ASEAN Neurological Association ; Start year: 2004
Added to DOAJ: 2005-10-03

29. **Signal Processing : An International Journal**
ISSN: 19852339
Subject: Computer Science
Publisher: Computer Science Journals ; Start year: 2007
Added to DOAJ: 2009-06-02

30. **UNITAR e-Journal**
ISSN: 15117219
Subject: Computer Science --- Social Sciences
Publisher: Universiti Tun Abdul Razak ; Start year: 2005
Added to DOAJ: 2005-08-25

31. **Wacana Seni Journal of Art Discourse**
ISSN: 16753410 ; EISSN: 19858418
Subject: Arts in general
Publisher: Universiti Sains Malaysia Press ; Start year: 2002
Added to DOAJ: 2010-04-26

Another pull factor to publish on Open Access is studies which indicate that OA journals are receiving citation and impact. A study by Testa and McVeigh (2004) (Table 4) who wanted to find out whether the OA journals performed differently from other non-OA journals in the various fields using ISI (Institute for Scientific Information) citation metrices.

In 2004, ISI covers about 200 OA journals and this number is small compared to the 8000 over journals indexed by the ISI then. They looked at a group of 148 journals in the natural sciences that have been covered long enough to have Impact Factors (IF) in the 2002 Journal Citation Reports (JCR). The results suggest that the OA journals have in general similar citation pattern to other journals, but may have a slight tendency to be cited earlier. The study found that there was a slightly higher percentage of citations to articles published in 2002. This situation is however discipline dependent. In fields such as pharmacology and mathematics, there is evidence of early citations. McVeigh (2004) found that in 2004 DOAJ, J-STAGE and Sci ELO listed a total of 1190 OA journal titles. Out of this number, 239 (20%) were indexed by the ISI which comprises 2.9% of the total 9000 titles indexed by the ISI. From January to June 2004, the number of OA journals had increased by 43 titles. The largest increase were in the fields of Physics, Engineering and Mathematics. Analysis of the performance of these OA journals show that the majority of OA journals are listed in the lower quartile category of journals in their subjects based on their impact factor. However, the OA journals performed better when ranked by their immediacy index. This means that because OA journals are made accessible earlier, the likelihood of being cited earlier is higher. This is especially true in the fields of medicine, life sciences, physics, engineering, mathematics and chemistry. The number of OA journals in the ISI databases have increased to 479 as reported in the JCR 2008 (revised version) (Giglia 2010; Hitchcock 2010; Agerback and Nielsen 2010). Also, 225 titles out of 479 (47%) show better performance in terms of the Immediacy Index than in Impact Factor (56% in Chemistry, 56% in Mathematics-Physics-Engineering, 41% in Life Sciences and 49% in Medicine). Eysenbach (2006), compares citation counts received by 1,492 articles (grouped into OA and non-OA journals) pubished in *PNAS: Proceedings of the National Academy of Sciences* between June and December 2004. Using a logistic regression model, he found that OA articles are more likely to receive citation than non OA articles and concluded that OA articles are likely to benefit through accelerated dissemination and early use.

Rowland and Nicholas (2005) reported on a study commissioned by the Publishers Association and the International Association of STM Publishers to find out the attitudes and perceptions of 5,513 authors about the new publishing models initiated by the digital environment. The authors were solicited from Australia, India, Mexico, France, Greece, the USA and UK, whose names were obtained from the mailing list of the Institute for Scientific Information (ISI). A total of over 76,000 email invitations to answer the questionnaire were posted and only 5,513 gave complete responses (7.2%). The respondents were active authors for they reported as being referees, journal editors or editorial board members in the previous 12 months. It was found that authors chose journals to publish (in order of priority) in terms of the following criteria;

- a)    reputation of the journal,
- b)    wide readership, journal with impact factor,
- c)    speed of publication,
- d)    reputation of the editorial board,
- e)    journal which allows preprint and post-print publishing,
- f)    as well as journals which allow authors to retain copyright.

The open ended sections revealed more information as authors indicate wanting the right to unlimited distribution and copyright of their work. Although the majority of authors felt that the reviewing process is important, many were dissatisfied with the time it takes. The authors in the sample also indicated high reliance on electronic medium to identify articles that are relevant to their needs. They follow links given in article references, use abstracting and indexing databases, search publishers' websites, search Google, Google Scholar and other search engines. The majority, (over 60%) have little knowledge or none at all about journals or institutional repositories. Authors anticipate the following outcomes of publishing (in order of importance):

a)      articles will be easier to obtain,
b)      libraries will have more money to spend,
c)      authors will publish more often,
d)      fewer articles will be rejected, and
e)      the quality of articles will improve.

A minority of the author (20.1%) thinks that OA publishing is a bad thing. A significant number of senior authors believe downloads to be a more credible measure of the usefulness of research. The results of this study indicate that there is a great deal that librarians can do to inform academics of publishing initiatives to make them more aware of their options to publish.

Table 4 : OA Journals Indexed by the ISI by Region (2004)

| Regions | No. of OA Journals | No. of Journals in ISI | Percent of OA |
|---------|---------|---------|---------|
| Asia-Pacific | 79 | 530 | 14.9% |
| Eastern Europe | 19 | 282 | 6.7% |
| Mid.East / Africa | 5 | 57 | 8.8% |
| North America | 58 | 3910 | 1.5% |
| South/Central America | 33 | 78 | 42.3 |
| Western Europe | 45 | 3961 | 1.1% |
| **WHOLE** | **239** | **8818** | **2.7%** |

Source: Testa and McVeigh (2004)

## Monitoring

Currently, repository's presence on the Web is being analysed by the Cybermetrics Laboratory in Spain (Ranking Web 2010; Aquillo 2010) which carries out quantitative studies about scientific communications through electronic journals and institutional repositories on the web. The ranking is done in accordance with the following indicators (Ranking Web 2010). Data from Open DOAR and ROAR was used for the analysis.

**Size (S)** = Number of pages recovered from the four largest engines: Google, Yahoo, Live Search and Exalead.
**Visibility (V) =** The total number of unique external links received (inlinks) by a site can be only confidently obtained from Yahoo Search and Exalead.
**Rich Files (R) =** The number of text files in Acrobat format (*.pdf*) extracted from Google and Yahoo.
**Scholar (Sc)** =  Calculate of the mean of the normalised total number of papers and those (recent papers) published between 2001 and 2008 found in Google scholar.

The four ranks were combined according to a formula where each one has a different weight but maintain the ratio 1:1 between activity (size *sensu lato*) and impact (visibility).

Based on the 2010 data, the Cybermetrics Lab listed the top 800 repositories. The ranking indicates that the top 20 repositories come from the United states, France, Germany and the European countries where establishing OA repositories are active(Table 5).

Table 5: Web Ranking of World Repositories (Top 20/800)

| World Rank | Repository | Country | Size | Visibility | Rich Files | Scholar |
|---|---|---|---|---|---|---|
| 1 | CiteSeerX | US | 2 | 1 | 528 | 2 |
| 2 | HAL Hyper Article en Ligne CNRS | FR | 9 | 5 | 1 | 7 |
| 3 | Research Papers in Economics | | 1 | 7 | 86 | 4 |
| 4 | Social Science Research Network | USA | 5 | 4 | 41 | 5 |
| 5 | Arxiv.org e-print Archive | USA | 19 | 2 | 231 | 3 |
| 6 | CERN Document Server | SWIS | 3 | 12 | 4 | 9 |
| 7 | Smithsonian/NASA Astrophysics Data System | USA | 11 | 3 | 739 | 1 |
| 8 | HAL Institut National de Recherche en Informatique et en Automatique Archive Ouverte | FR | 10 | 11 | 5 | 21 |
| 9 | Digital Lib and Archives Virginia Tech University | USA | 13 | 10 | 3 | 33 |
| 10 | HAL Hyper Article en Ligne Sciences de l'Homme et de la Société | FR | 16 | 9 | 7 | 39 |
| 11 | École Poly. Federale de Lausanne Infoscience | SWIS | 4 | 13 | 11 | 137 |
| 12 | MIT DSpace | USA | 15 | 27 | 6 | 11 |
| 13 | Ressources documentaires Institut de recherche pour le développement | FR | 8 | 23 | 2 | 304 |
| 14 | Calif Inst of Tech Online Archive of California | USA | 7 | 15 | 8 | 683 |
| 15 | Depot Erudit | CA | 119 | 8 | 153 | 347 |
| 16 | Organic ePrints | DE | 22 | 38 | 22 | 30 |
| 17 | Univ of Southhampton Dept Elec. & Comp. Sci | UK | 24 | 22 | 37 | 98 |
| 18 | Humbolt Universitat zu Berlin Publikationsserver | GER | 26 | 30 | 24 | 123 |
| 19 | Tufts University Perseus Digital Library | USA | 6 | 6 | 477 | 809 |
| 20 | Universitat Stuttgart Elektronische Hochschulschriften | GER | 77 | 14 | 43 | 292 |

Source: Ranking Web.. Cybermetrics, Lab, Spain, 10 July 2009 http://repositories.webometrics.info/top800_rep.asp

Table 6 indicates that amongst the top 20 out of 100 OA initiatives listed under Southeast Asian countries, the active countries are Thailand, followed by Malaysia, Indonesia and Singapore. Amongst the 800 repositories, the repositories at UPM was ranked at 159, UUM at 246, UM at 356, UKM at 408 and Universiti Telekom Petronas at 559.  The performance of Malaysian universities in terms of Southeast Asian countries is indicated in Table 6. It is curious to note that except for those top ranked universities which are active in OAI research (MIT, Virginia Tech in the USA), the other highly ranked universities seem less active. The *Oxford University E print repository* is ranked at 504 and the *Oxford University Research Archive* is ranked at 707. The performance of Malaysian repositories amongst the 8000 world repositories is given in Table 7.

The Cybermetrics Laboratory also provides ranking by country (Table 8) using a different sets of indicators listed below:

- **System**: Number of universities in the Top 500 in the given country, divided by the mean position of those institutions.
- **Access**: A score built according to ranks (5 points for a university in the top 100, 4 points for 101-200, 3 points for 201-300, 2 for 301-400 and 1 for 401-500) divided by the population size (root of the population in thousands) of the country (World Bank, 2007).
- **Flagship**: A normalized score (100 for positions 1-20, 96 for 21-40, and so on) based on the leading university rank for countries with institutions among the Top 500.
- **Economic**: Same score as the access defined before but divided by the GDP (PPP) per capita for the country in question (World Bank, 2007).

The advice given by the Cybermetrics lab group is as follows.
" If the web performance of an institution is below the expected position according to their academic excellence, institution authorities should reconsider their web policy, promoting substantial increases of the volume and quality of their electronic publications".

Table 6: Web Ranking of Universities in Southeast Asia (Top 20/100)

| Rank SEA | University | Country | World Rank | Size | Visibility | Rich Files | Scholar |
|---|---|---|---|---|---|---|---|
| 1 | National University of Singapore | SG | 146 | 120 | 210 | 122 | 192 |
| 2 | Kasetsart University | TH | 229 | 459 | 156 | 324 | 354 |
| 3 | Prince of Songkla University | TH | 338 | 236 | 236 | 658 | 587 |
| 4 | Mahidol University | TH | 381 | 473 | 394 | 947 | 91 |
| 5 | Chulalongkorn University | TH | 398 | 541 | 445 | 474 | 291 |
| 6 | Nanyang Technological University | SG | 468 | 434 | 560 | 718 | 311 |
| 7 | Chiang Mai University | TH | 478 | 666 | 414 | 523 | 807 |
| 8 | Universitas Gahjah Mada | IND | 562 | 602 | 421 | 1,028 | 827 |
| 9 | Khon Kaen University | TH | 567 | 824 | 387 | 703 | 1,214 |
| 10 | Institut Teknologi Bandung | IND | 661 | 564 | 657 | 1,138 | 654 |
| 11 | Universiti Putra Malaysia | MAL | 686 | 688 | 887 | 1,064 | 342 |
| 12 | Thammasat University | TH | 700 | 525 | 748 | 1,035 | 846 |
| 13 | Universiti Sains Malaysia | MAL | 725 | 500 | 1,097 | 1,195 | 256 |
| 14 | Universiti Teknologi Malaysia | MAL | 733 | 519 | 1,487 | 852 | 126 |
| 15 | Asian Institute of Technology Thailand | TH | 770 | 436 | 846 | 1,218 | 992 |
| 16 | University of Malaya | MAL | 778 | 857 | 1,1328 | 1,239 | 100 |
| 17 | University of Indonesia | IND | 815 | 903 | 1,007 | 741 | 981 |
| 18 | King Mongkut University of Technology | TH | 822 | 836 | 682 | 1,100 | 1,437 |
| 19 | Petra Christian University | TH | 854 | 1137 | 1794 | 964 | 59 |
| 20 | Naresuan University | TH | 924 | 1687 | 591 | 1088 | 1775 |
| REST | | | | | | | |
| 22 | Universiti Kebangsaan Malaysia | MAL | 985 | 849 | 1515 | 1063 | 614 |
| 35 | Universiti Teknologi Mara | MAL | 1,367 | 1010 | 1285 | 1870 | 2254 |
| 36 | Universiti Malaysia Perlis | MAL | 1413 | 1697 | 1518 | 3974 | 501 |
| 37 | Universiti Utara Malaysia | MAL | 1454 | 1623 | 2249 | 1531 | 776 |
| 39 | Multimedia University | MAL | 1528 | 1173 | 1595 | 1771 | 2250 |
| 42 | International Islamic University | MAL | 1576 | 2199 | 1510 | 2034 | 1747 |
| 66 | University of Nottingham Malaysia | MAL | 2273 | 5233 | 2256 | 3927 | |
| 78 | Universiti Malaysia Pahang | MAL | 2546 | 3280 | 4136 | 1341 | 2088 |
| 83 | Universiti Tenaga Nasional | MAL | 2665 | 2877 | 4348 | 2244 | 1632 |
| 84 | Universiti Malaysia Sabah | MAL | 2681 | 2271 | 2844 | 4153 | 2899 |

Source: Ranking Web, Cybermetrics Lab, Spain , 10 July 2010. Available at
http://www.webometrics.info/top100_continent.asp?cont=SE_Asia

Table 7: Performance of Malaysian Universities Repositories

| World Rank | University | Size | Visibility | Rich files | Scholar |
|---|---|---|---|---|---|
| 686 | Universiti Putra Malaysia | 688 | 887 | 1064 | 342 |
| 725 | Universiti Sains Malaysia | 500 | 1097 | 1195 | 256 |
| 733 | Universiti Teknologi Malaysia | 519 | 1487 | 852 | 126 |
| 778 | University of Malaya | 857 | 1328 | 1239 | 100 |
| 985 | Universiti Kebangsaan Malaysia | 849 | 1515 | 1063 | 614 |
| 1367 | Universiti Teknologi Mara | 1010 | 1285 | 1870 | 2254 |
| 1413 | Universiti Malaysia Perlis | 1697 | 1518 | 3974 | 501 |
| 1454 | Universiti Utara Malaysia | 1623 | 2249 | 1531 | 776 |
| 1528 | Multimedia Universiti | 1173 | 1595 | 1771 | 2250 |
| 1576 | International Islamic University Malaysia | 2199 | 1510 | 2034 | 1747 |
| 2273 | University of Nottingham Malaysia | 5233 | 2256 | 3927 | 1091 |
| 2546 | Universiti Malaysia Pahang | 3280 | 4136 | 1341 | 2088 |

| 2665 | Universiti Tenaga Nasional | 2877 | 4348 | 2244 | 1632 |
|------|---------------------------|------|------|------|------|
| 2681 | Universiti Malaysia Sabah | 2271 | 2844 | 4153 | 2899 |
| 3185 | Universiti Teknologi Petronas | 5571 | 1704 | 7812 | 3848 |
| 3212 | Open University Malaysia | 4022 | 3388 | 4632 | 2776 |
| 3222 | Monash University Malaysia | 5186 | 2446 | 4069 | 4539 |
| 3441 | Universiti Sarawak Malaysia | 4622 | 2116 | 6946 | 4857 |
| 3508 | Universiti Tun Hussein Onn Malaysia | 2446 | 5039 | 2524 | 4053 |
| 3741 | Universiti Malaysia Terengganu | 4914 | 4174 | 4889 | 2801 |
| 3788 | Curtin University of Technology Sarawak Campus | 7085 | 2029 | 5726 | 6624 |
| 3874 | Universiti Pendidikan Sultan Idris | 2769 | 3424 | 5014 | 7086 |
| 4060 | Taylor's University College | 5031 | 5360 | 6052 | 1876 |
| 4149 | Islamic Science University of Malaysia | 3870 | 5595 | 7385 | 1831 |
| 4165 | Universiti Tun Abdul Razak | 3439 | 5365 | 5084 | 3321 |
| 4197 | Universiti Teknikal Malaysia Melaka | 3099 | 5361 | 3674 | 4893 |
| 5065 | University of Kuala Lumpur | 6128 | 3008 | 9389 | 8563 |
| 5276 | Sunway University College | 5968 | 6114 | 7232 | 3287 |
| 5413 | Universiti Tunku Abdul Rahman | 5540 | 6198 | 4541 | 5946 |
| 5763 | Asia Pacific Institute of Information Technology | 4643 | 5656 | 6722 | 6001 |
| 5834 | Wawasan Open University | 6644 | 6441 | 7093 | 4187 |
| 6104 | Universiti Darul Iman Malaysia | 5267 | 5186 | 8572 | 8563 |
| 6353 | Universiti Industri Selangor | 5968 | 5126 | 9356 | 8563 |
| 6358 | UCSI University | 8026 | 6816 | 5408 | 5547 |
| 6395 | Help University College | 5060 | 6498 | 7353 | 7631 |
| 6662 | Tunku Abdul Rahman College | 7882 | 5182 | 11331 | 6004 |
| 6788 | KDU College | 5352 | 5269 | 11453 | 9750 |
| 6884 | University of Malaya Medical Centre & Faculty of Medicine | 7439 | 6546 | 8452 | 6098 |
| 7077 | Malaysia Theological Seminary | 11580 | 4471 | 8348 | 8563 |
| 7089 | Universiti Malaysia Kelantan | 7858 | 5877 | 9378 | 8001 |
| 7193 | Segi College | 7387 | 7007 | 7791 | 7252 |
| 7235 | International Medical University | 8596 | 7052 | 9764 | 4715 |
| 7429 | LimKokWing University of Creative Technology | 6068 | 5852 | 12727 | 9750 |
| 7907 | National Defence University of Malaysia | 10537 | 4916 | 12406 | 9750 |

Source: Ranking Web Universities by countries, Cybermetrics lab, 10 July 2010. Available at:
http://www.webometrics.info/rank_by_country.asp?country=my

Table 8: Distribution of Repositories by Continent

| Continent | Top 200 | Top 500 | Top 1000 |
|-----------|---------|---------|----------|
| USA & Canada | 114 | 200 | 370 |
| Europe | 60 | 223 | 408 |
| Asia | 15 | 45 | 124 |
| Oceania | 6 | 14 | 35 |
| Latin America | 4 | 14 | 44 |
| Arab World | 1 | 5 | 4 |
| Africa | | 1 | 5 |

Source: Ranking Web, Cybermetrics Lab, Spain , 10 July 2010.
Available at http://www.webometrics.info/Distribution_by_Country.asp

## IMPLICATIONS FOR LIBRARIES

### Aware of OA Issues

Open Access repositories should have an impact on libraries, especially academic libraries. First and foremost, librarians must be knowledgeable about what OA means, the differences between OA and free access, what is open repositories, creative commons license, e-prints, post-prints, self archiving, OA journals, how do users search for OA documents, how OA repositories and journals affect the library's collections and in Malaysia how affects institutional research visibility and impact. Bailey (2006) proposed that librarians should be able to advice university management about the feasibility of setting up institutional repositories using OA software to increase institutional visibility and impact and to educate academics about self-archiving, to inform about the types of institutional materials that can be archived and the degree of accessibility given to users. The presence of Malaysian academic libraries in ROAR shows an awareness of academic librarians about this issue. However, the spirit of OA is not being readily assimilated as restrictions are being imposed by some libraries, to their academic's journal articles and especially theses and dissertations. should mean "removing permission barriers". Libraries may adopt the Creative Commons license agreement where the rights remain with the authors who may grant users with certain rights or the authors give up all rights and makes his work available freely to the public.

Libraries could also embed QA issues into user education programmes at both undergraduate, postgraduate and academic staff levels. This would help increase awareness and adoption.

**Remove Price Crisis and Limited Permission Crisis**

Suber (2003) proposed that advocating to help remove all woes faced by libraries as it removes serials pricing crises, remove legal barriers from copyright laws and license agreement. He observed that even though libraries pay huge sums of money to subscribe journals, their freedom to archive is limited by licensing agreements. In a sense, libraries now pay more "in order to get much less". Suber terms the crises libraries face as "permission crisis". Adopting the policy makes scholarly literature become available to everyone and users are allowed to read, download, share, store, print, link and cite. Just think if all librarians manage to influence the academic publishers in their institutions to adopt the OA policy, there would be no more subscriptions, pricing issues are solved, permission crisis is removed and what exist are mutual linking of inter-university repositories, a situation of sharing and using. The costs of providing OA repositories and OA journals are absorbed by the funding institutions. But of course this is an ideal situation as there will always be those who are overly cautious about opening access to their institutional scholarly works which are usually underuse because other users are simply unaware of their existence.

Suber (2004) identified the advantages libraries get when advocating OA repositories and OA journals:
- Libraries have the right to archive for example, journal issues as a backup to existing sites or to archive past print issues to supplement those available online. This is especially true for journals which have long print runs.
- Libraries would be able to convert materials to new media format to keep them readable as technology changed.
- Libraries would make all materials available to all users, on and off site.
- Users would not be limited by password, IP address, usage hours or ability to pay.
- Libraries could emphasize that faculties should donate their research papers to the repositories to increase visibility. In return, faculty could equally use other items available in the repositories.
- Libraries do not have to negotiate for prices or licensing terms.
- Libraries need not cancel serials due to reduced budget.

**Help in the Development of Institutional Repositories**

Bailey (2006), itemized how libraries can help in the development of institutional repositories :

- Help create IR policies and procedures
- Provide feedback about the work process of the IR
- Assist in designing the IR user interface so that it is clear, easy to use, and effective
- Help identify self-archiving activity and processes in their various institutions
- Promote the IR to faculty and graduates
- Introduce IR in user education programmes
- Provide assistance to academics and students on how to deposit items and search for items.
- Help in enhancing the descriptors or metadata to items in IRs
- Help in monitoring the quality control of deposits in IRs as chief cataloguers have done for years for library catalogue entries.

**Provide Advice to AO Journal Publishers**

Librarians can advice institutional academic journal publishers to adopt the golden publishing model used by *PLoS Medicine*, if institutional support is not available. Ask authors to pay a minimal sum to publish their refereed and accepted article and users are given full access. This situation is quite plausible as most researchers obtain funding for their research and the cost of publishing can be absorbed by such funds.

**Libraries as Publishers of OA journals**

Libraries have been involved in publishing electronic journals. The University of Houston published *The Public-Access Computer Systems Review* in 1989. In 1990s, the Scholarly Communications Project of The Virginia Polytechnic Institute published the *Journal of the International Academy of Hospitality Research*.

**Libraries Can Collaborate in Building Special Collections**

There have been successful OA initiatives which involve the collaboration between both libraries and research groups. Examples are MIT libraries/Hewlett Packard in *DSpace* (Mackenzie, et al. 2003) and University of Virginia Libraries in *Fedora* (Staples, Wayland and Payette, 2003). At the University of Malaya, two initiatives grew from such collaboration. Libraries often synchronize this collaborative venture as part of their digitization project. This is absolutely true for special libraries which are not well funded. The *Dewan Bahasa dan Pustaka* Library (DBPL) has benefitted in collaborating with the University of Malaya Digital Library Research Group in acting as the content expert and provider for the digital library of Malay manuscripts (*MyManuskrip*), funded by the Ministry of Science, Technology and Innovation between 2007 and 2009 (Figure 3).

In this collaborative repository (Zainab, Abrizah and Hilmi 2009), DBPL has successfully digitized 69 titles of original Malay manuscript costing close to RM60,000 using allocations from the research grant. Another partner of the digital library is the University of Malaya Library which benefitted from the project by getting 102 titles of their Malay manuscripts digitized which cost about RM90,000 and a dedicated microform scanner used to convert microform version of manuscripts to the digital format (costing over RM50,000). *MyManuskrip* is listed in ROAR as a cross-institutional repository and currently holds about 179 digital Malay manuscripts.

Another collaborative effort is *DSpace@UM* (Abrizah 2009), an institutional repository that provide access to over 763 theses and dissertation mainly submitted to the University of Malaya. The UM library coordinates the collection of digital copies of students theses submissions and provide them to the Digital Library Research Group to be used the content for simulation. This project is funded by the University of Malaya Research Grant between 2009 and 2010 (Figure 4).

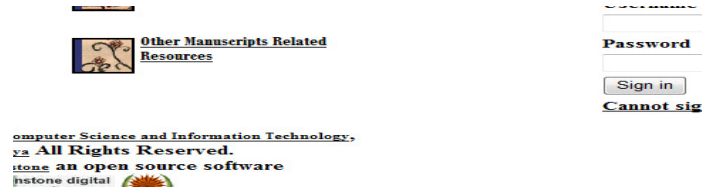

Figure 3: Main Page of MyManuskrip.Available at
http://mymanuskrip.fsktm.um.edu.my/Greenstone/cgi-bin/library.exe

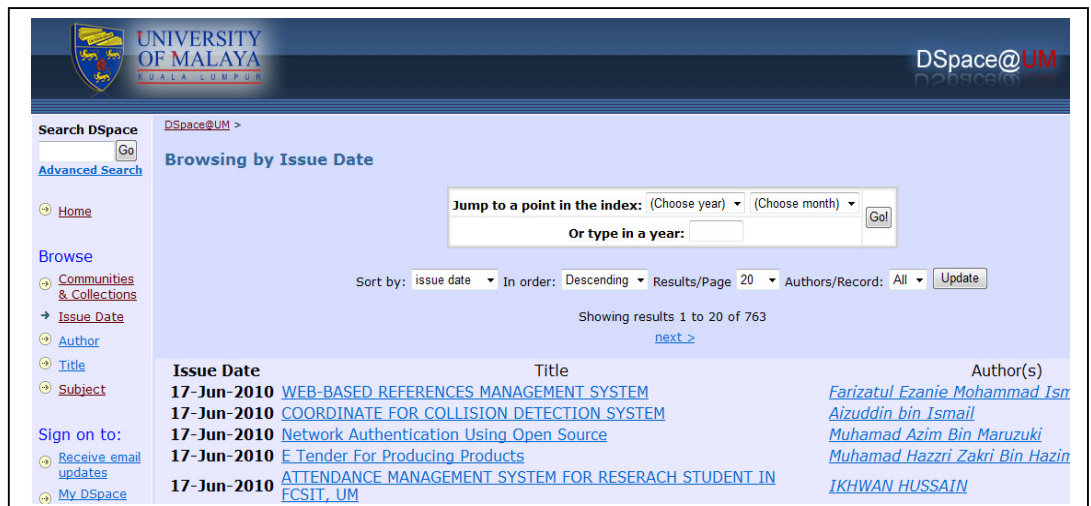

Figure 4: Main Page of DSpace@UM

Abrizah (2009) assessed the readiness of University of Malaya academics in accepting and contributing to *D Space @UM*. She surveyed 131 academics from 14 faculties and reported favourable response from the science-based faculty members. About 60% of the respondents were in favour of depositing theses and dissertations in the repository.

Respondents were motivated by the principle and understood that this would make their work more accessible and visible. There was fear about copyright and plagiarism issues and the fear that depositing their pre-prints would prevent their work from getting published. The respondents also perceived that a mandate from the university and funding bodies would help put these worries to rest.

In both these collaborative initiatives, libraries win in terms of;

- Getting their rare items digitized without incurring any financial cost
- Obtaining more experience in setting up an OA repository and understanding the processes that need to be structured; and
- Making their institutional research and rare items more visible and open to the public to be read, use and cited.

Van Westrienen and Lynch (2005) clearly say it all. Their world survey of institutional repositories revealed problems such as the difficulties of convincing faculties about the value of institutional repositories; problems of convincing them to contribute their works, problem of ironing out issues of copyright and intellectual property, the suspicion authors have with rights, the problem of dispelling believes of losing impact and scholarly credit, and the cumbersome submission system of some archives which "put off" some faculties. Problems such as copyright ownerships seem to be able to be resolved, as attitudes are beginning to change. More e-print, e-journal and print journal publishers are going green by allowing authors to retain the copyright to their works. The *Electronic Journal of Comparative Law* (*EJCL*), *British Medical Journal* and *Nucleic Acid Review* (the latter two had recently switched to an open source model) allow their authors to keep their copyright and this is mentioned in the copyright notice that articles to be produced for educational purposes and other uses should seek author's permission. In this situation, the publisher asks only for a license to publish the article as the first publisher. Authors are allowed to republish their article on other platforms and are obliged to mention *EJCL* as the original source. This is becoming a typical copyright policy adopted by many electronic journals and the majority of authors (71%) who published in journals also agree that they should be allowed to keep copyright of their works (Hoon 2006).

## CONCLUSION

OA electronic journals, e-print repositories and archives could make Malaysian research available and visible and increase the chances for use and exchange of ideas among scholars within similar disciplines. The "end" of scholarly communication may therefore be fulfilled, that is to provide an environment for scholarly inter-communication, establishing recognition for authors, conferring authors with the right to disseminate various versions of their articles as reflected by an on-going research activity, and allowing authors to disseminate to the largest audience possible. The future of scholarly communication will definitely be dominated by OA electronic journals and archives as a channel for communication, and should be planned on an initiative in various focused subject areas as exemplified by *arXiv.org* and *E-print in Library and Information Science (E-LIS)*, which encourage authors to submit their articles to the e-print repositories.

Where do libraries fit in this situation? Subject librarians and faculty liaison librarians could play the role of creating awareness amongst academics of the various faculties they are responsible for. To contend that academics know about the existence of electronic journals and repositories within their discipline is more often a fallacy. With the current universities' emphasis for their faculty members to publish in journals, librarians could help by making them aware that more OA journals are refereed, some are highly cited and indexed by the ISI databases and SCOPUS. With a little bit of homework, the data can be given to them as evidence. Academics could be informed either through personal emails or an online directory of electronic journals categorized by broad disciplines as reflected by the faculties which exist within the university. Subject indexes could inform academics of the types of OA journals, and e-print archives that are available in the respective disciplines, and also provide information such as the refereeing status of the journals, their impact factor, if any, and whether they are on Open Access. On top of this, the OA journals and repositories electronic journals should be catalogued as a resource, searchable in the library's OPAC and actively linked to the actual electronic OA journals. Ultimately, it is the case of, whether the academics' knowledge about the OA electronic journals will induce them to start using it for dissemination and research consumption - a case of "to know is to use".